\documentclass[11pt, a4paper]{article}

\usepackage[nonatbib, final]{neurips_2023}

\usepackage[utf8]{inputenc} 
\usepackage[T1]{fontenc}    
\usepackage{hyperref}       
\usepackage{url}            
\usepackage{booktabs}       
\usepackage{amsfonts}       
\usepackage{nicefrac}       
\usepackage{microtype}      
\usepackage{xcolor}         
\usepackage{bbm}
\usepackage{amsthm}
\usepackage{rotating}

\usepackage{graphicx} 
\usepackage[ngerman, english]{babel}
\usepackage[iso, ngerman]{isodate}

\usepackage{mathtools}
\usepackage{amsmath} 
\usepackage{nicefrac}
\usepackage{tikz}
\usepackage{subcaption}
\usepackage{centernot} 

\newtheorem{theorem}{Theorem}

\usepackage[colorinlistoftodos]{todonotes}
\usepackage{verbatim}

\usepackage{natbib} 
\DeclareRobustCommand{\independent}{\mathbin{\perp\!\!\!\perp}}

\newcommand{\nprobInd}[1]{%
    \mathrel{\centernot{\mkern-3.5mu\mathrel{#1}}}
}

\newcommand{\Ppost}{P_{\text{post}}}
\newcommand{\Ppre}{P_{\text{pre}}}

\title{Performativity and Prospective Fairness}
\author{%
  Sebastian Zezulka
  \\
  University of Tübingen\\
  \texttt{sebastian.zezulka@uni-tuebingen.de} \\
   \And
  Konstantin Genin \\
  University of Tübingen \\
  \texttt{konstantin.genin@uni-tuebingen.de} \\
}

\begin{document}

\maketitle

\begin{abstract}
    Deploying an algorithmically informed policy is a significant intervention into the structure of society. As is increasingly acknowledged, predictive algorithms have performative effects: using them can shift the distribution of outcomes away from the one on which the algorithms were trained. Algorithmic fairness research is usually {\em motivated} by the worry that these performative effects will exacerbate the structural inequalities reflected in the training data. However, standard {\em retrospective} fairness methodologies are ill-suited to account for such performativity. They impose static fairness constraints that hold after the predictive algorithm is trained, but before it is deployed and, therefore, before performative effects have had a chance to kick in. Unfortunately, satisfying static fairness criteria after training is not sufficient to avoid exacerbating inequality after deployment. Our first contribution is conceptual: we argue that addressing the fundamental worry that motivates algorithmic fairness requires a notion of {\em prospective} fairness that anticipates the change in relevant structural inequalities \textit{after} deployment. Our second contribution is methodological: we propose a strategy for estimating this post-deployment change from pre-deployment data. That requires distinguishing between, and accounting for, different kinds of performative effects. In particular, we focus on the way predictions change policy decisions and, therefore, the distribution of outcomes. Throughout, we are guided by an application from public administration: the use of algorithms to (1) predict who among the recently unemployed will remain unemployed in the long term and (2) targeting them with labor market programs. We illustrate our proposal by showing how to predict whether such policies will exacerbate gender inequalities in the labor market.
\end{abstract}

\section{A fundamental question for fair machine learning}
\label{sec:fundamental}
Research in algorithmic fairness is usually motivated by the worry that machine learning algorithms will reproduce or exacerbate the structural inequalities reflected in their training data \citep{lum2016predict, Tolbert2023}. Indeed, whether an algorithm exacerbates an existing social inequality is emerging as a central compliance criterion in EU non-discrimination law \citep{Weerts2023_CONF}. However, the methodological solutions developed by researchers in algorithmic fairness are, surprisingly, ill-suited for answering this fundamental question. In order to decide whether embedding some algorithm into our socio-technical processes exacerbates existing inequalities, we must make some effort to, first, identify the contextually relevant inequalities and, second, predict whether the new algorithmic policy will exacerbate them. Most algorithmic fairness methods are \textit{retrospective} insofar as they usually do not attempt the latter. Moreover, by failing to have this latter goal in mind, they typically also struggle to identify the relevant inequalities in real world outcomes, focusing instead on internal features of the algorithm. Since most structural inequalities long predate prediction algorithms, internal fairness properties of the algorithm are, at best, a proxy for the relevant inequalities. Our contributions are as follows: 
\begin{enumerate}
    \item We {\em reconceptualize} fairness questions as policy problems. {\em Prospective fairness} is a matter of predicting whether deploying an algorithmically informed policy will exacerbate inequality.
    \item We propose a {\em methodology} to identify the effect of deploying an algorithmically informed policy on context-relevant inequalities from pre-deployment data.
    \item We illustrate the proposal by a case study in the statistical profiling of registered unemployed. For a concrete example, we study the likely effects of two algorithmic policy proposals on the gender gap in long-term unemployment rates.
\end{enumerate}

The plan of the paper is as follows: first, we argue for {\em prospective fairness} as a conceptual framework and survey related work; section~\ref{sec:statprofiling} introduces two recently proposed algorithmic policies intended to support public employment agencies in reducing long-term unemployment; we argue that, in this context, the dependence between gender and employment outcome, as well as the gender gap in reemployment probabilities, are simple and intuitive measures of systemic inequality; section~\ref{sec:perfandpros} enumerates the challenges posed by different performative effects for predicting a post-deployment measure of systemic inequality from pre-deployment data and proposes a method for overcoming (some of) them; section~\ref{sec:toymodel} illustrates the method with a simple toy model and section~\ref{sec:Conclusion} outlines directions for future work.

\subsection{From retrospective to prospective fairness}
In paradigmatic risk-assessment applications, machine learners are concerned with learning a function that takes as input some features $X$ and a sensitive attribute $A$ and outputs a score $R$ which is valuable for predicting an outcome $Y$. The algorithmic score $R$ is meant to inform some important decision $D$ that, typically, is causally relevant for the outcome $Y$. In the application that concerns us in this paper, features such as the education and employment history $(X)$ and gender $(A)$ of a recently unemployed person are used to compute a risk score $(R)$ of long-term unemployment $(Y).$ This risk score $R$ is meant to support a case-worker at a public employment agency in making a plan $(D)$ about how to re-enter employment. This plan may be as simple as requiring the client to apply to some minimum number of jobs every month or referring them to one of a variety of job-training programs. 

Formal fairness proposals require that some property is satisfied by either the joint distribution $P(A,X,R,D,Y)$ or the causal structure $G$ giving rise to it. Individual fairness proposals introduce a similarity metric $M$ on $(A,X)$ and suggest that similar individuals should have similar risk scores. In all these cases, the relevant fairness property is a function $\varphi(P,G,M)$. Group-based fairness \citep{barocas-hardt-narayanan} ignores all but the first parameter; causal fairness \citep{kilbertus2017avoiding, kusner2017counterfactual} ignores the last; and individual fairness  \citep{Dwork2012_CONF} ignores the second. All these proposals agree that fairness is a function of the distribution (and perhaps the causal structure) at the time when the prediction algorithm has been trained, {\em but before it has been deployed}. Our first contribution is to argue that addressing the fundamental question of fair machine learning requires comparing the status quo {\em before} deployment with the situation likely to arise {\em after} deployment. In other words: {\em prospective} fairness is a matter of anticipating the change from $\varphi(P_{\text{pre}}, D_{\text{pre}}, M)$ to $\varphi(P_{\text{post}},D_{\text{post}}, M)$. We do not claim that there is a single correct inequality measure $\varphi(\cdot),$ nor even that there is an all-things-considered way of trading off different candidates, only that we must make a good faith effort to anticipate changes in the relevant measures of inequality. 

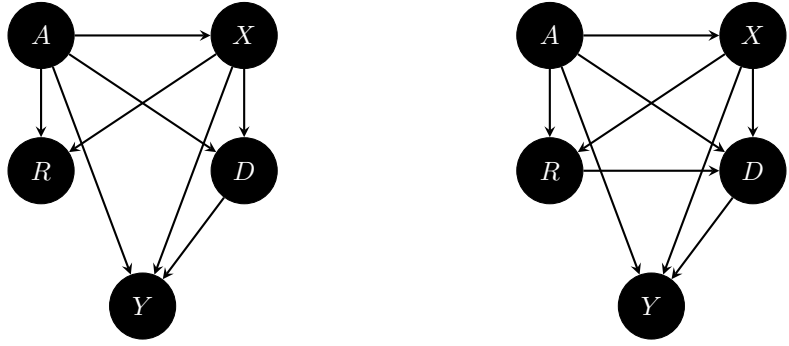
\begin{figure}[h]
  \centering
  \begin{subfigure}{0.4\textwidth}
    \centering
    \begin{tikzpicture}[>=stealth, every node/.style={circle, draw, fill=black, text=white, minimum size=25pt}, scale=0.9]
          \node (A) at (0,0) {$A$};
          \node (X) at (3,0) {$X$};
          \node (R) at (0,-2) {$R$};
          \node (D) at (3,-2) {$D$};
          \node (Y) at (1.5,-4) {$Y$};
        
          \draw[->, thick] (A) -- (R);
          \draw[->, thick] (A) -- (X);
          \draw[->, thick] (A) -- (D);
          \draw[->, thick] (A) -- (Y);
          
          \draw[->, thick] (X) -- (R);
          \draw[->, thick] (X) -- (D);
          \draw[->, thick] (X) -- (Y);
          
          \draw[->, thick] (D) -- (Y);
    \end{tikzpicture}
    \caption{Causal structure $G_{\text{pre}}$ \textit{before} deploying an algorithmically-informed policy.}
  \end{subfigure}
  \hspace{1cm}
  \begin{subfigure}{0.4\textwidth}
    \centering
    \begin{tikzpicture}[>=stealth, every node/.style={circle, draw, fill=black, text=white, minimum size=25pt}, scale=0.9]
        \node (A) at (0,0) {$A$};
        \node (X) at (3,0) {$X$};
        \node (R) at (0,-2) {$R$};
        \node (D) at (3,-2) {$D$};
        \node (Y) at (1.5,-4) {$Y$};
    
        \draw[->, thick] (A) -- (R);
        \draw[->, thick] (A) -- (X);
        \draw[->, thick] (A) -- (D);
        \draw[->, thick] (A) -- (Y);
        
        \draw[->, thick] (X) -- (R);
        \draw[->, thick] (X) -- (D);
        \draw[->, thick] (X) -- (Y);
        
        \draw[->, thick] (R) -- (D);
        \draw[->, thick, thick] (D) -- (Y);
    \end{tikzpicture}
    \caption{Causal structure $G_{\text{post}}$ \textit{after} deploying an algorithmically-informed policy.}
  \end{subfigure}
    \caption{The left hand side shows the pre-deployment causal graph $G_{\text{pre}}$ inducing a joint probability distribution $P_{\text{pre}}$ over sensitive attributes $A$, features $X$, risk score $R$, decision $D$, and outcome variable $Y$. The  risk score $R$ is the output of a learned function from $A$ and $X$. Since this graph represents the situation after training, but before deployment, there is no arrow from the risk score $R$ to the decision $D$. \textit{Retrospective} fairness formulates constraints $\varphi(G_{\text{pre}}, P_{\text{pre}}, M)$ on the pre-deployment arrangement alone. The right-hand side represents the situation after the algorithmically informed policy has been deployed, with predictions $R$ now affecting decisions $D$. Prospective fairness requires comparing the consequences of intervening on the structure of $G_{\text{pre}}$ and moving to $G_{\text{post}}$. In other words, comparing $\varphi(G_{\text{pre}}, P_{\text{pre}}, M)$ with $\varphi(G_{\text{post}}, P_{\text{post}}, M)$.}
  \label{fig:DAG_D0_D1}
\end{figure}

 As shown in Figure~\ref{fig:DAG_D0_D1}, deploying a decision support algorithm introduces a causal path from the predicted risk score $R$ to the decision $D$. Importantly, the outcome variable $Y$ is causally downstream of this intervention. The addition of a causal path is a {\em structural} intervention \citep{malinsky2018intervening,Bynum2022} not captured by standard surgical $do(X=x)$ interventions that remove edges and fix a variable to a specific value.

From a dynamical perspective, static and retrospective fairness proposals go wrong in two ways. In the worst case, they are {\em self-undermining}: satisfying the fairness criteria at the time of training necessitates violating them after implementation. For example, \citet{Mishler2022} show that satisfying the fairness notions of sufficiency $(Y\independent A~|~R)$ or separation $(R\independent A~|~Y)$ at the time of training virtually ensures that they will be violated after deployment. Illustrating the point in terms of sufficiency, where $\independent$ denotes (conditional) statistical independence:  
\[ Y\independent_{\text{pre}} A~|~R \hspace{4pt}\text{ entails } \hspace{4pt}  
Y \nprobInd{\independent}_{\text{post}} A~|~R.  \] 

Group-based notions of fairness like sufficiency and separation fall victim to {\em performativity}: the tendency of an algorithmic policy intervention to shift the distribution away from the one on which it was trained \citep{perdomo2020performative}. But as \citet{Mishler2022} show, they are undermined not by an unintended and unforeseen performative effect, but by the {\em intended, and foreseen} shift in distribution induced by algorithmic support, i.e.:
\[P_{\text{pre}}(D~|~A,X,R)\neq P_{\text{post}}(D~|~A,X,R).~\]
In other words, they are undermined by the fact that algorithmic support changes decision-making, which, presumably, is the point of algorithmic support in the first place. Since the distribution of the outcome $Y$ will change after deployment, \citet{berk2021improving} advises against group-based metrics involving it, opting for simple independence ($R\independent A$) instead. Of course, independence requires a loss of predictive accuracy, which may undermine even the most benevolent policies.

It is not likely that individual and causal fairness proposals are so drastically self-undermining. So long as the similarity metric stays constant, an algorithm that treats similar people similarly will continue to do so after deployment. If, as \citet{kilbertus2017avoiding} suggest, causal fairness is a matter of making sure that all paths from the sensitive attribute $A$ to the  prediction $R$ are appropriately mediated, then causal fairness is safe from performative effects so long as the qualitative causal structure \textit{upstream} of the prediction $R$ remains constant.  But even if causal and individual fairness proposals are not so dramatically self-undermining, they are simply {\em not probative} of whether the algorithm reproduces or exacerbates existing inequalities since these effects are causally \textit{downstream} of algorithmic predictions. In particular, it is customary to ignore the real-world dependence between $A$ and $Y$ induced by the social status quo as target of an intervention, since nothing can be done about it at the time of training. Instead, fairness researchers focused on whether the risk score \textit{itself} is fair, whether in the group, individual or causal sense. However, from the dynamical perspective, it is perfectly reasonable to ask whether the proposed algorithmic policy will exacerbate the systemic inequality reflected in the dependence between gender $(A)$ and long-term unemployment $(Y)$. Indeed, simple dynamical models and simulations suggest that algorithms meeting static fairness notions at training may exacerbate inequalities in outcomes in the long-run \citep{Liu2019_CONF,Zhang2021_CONF}. Streamlined dynamical models and simulations are a valuable tool in evaluating the long-run effects of fairness-constrained algorithms. The second contribution of this paper, however, is a methodology for estimating the systemic effect of a proposed algorithmic policy from pre-deployment data. Of course, we would not expect such a procedure to exist in all cases. Rather, we hope to show that this is possible under not-too-heroic assumptions.

\subsection{Related Work}\label{sec:relatedwork}
In machine learning, the fairness debate began with risk assessment tools for decision- and policy-making \citep{Angwin2016, Kleinberg2016, Chouldechova2017, Mitchell2021}. To this day, many standard case studies e.g., lending, school admissions, and pretrial detention, fall within this scope. See \citet{Berk2023} for a review on fairness in risk assessment and \citet{Borsboom2008} and \citet{Hutchinson2019_CONF} for predecessors in psychometrics. Since then, researchers have stressed the importance of explicitly differentiating policy decisions from the risk predictions that inform them \citep{Barabas2018_CONF, Kuppler2021, Beigang2022} and of studying machine learning algorithms in their socio-technological contexts \citep{Selbst2019_CONF}. We incorporate both of these insights into the present work.

A central negative result emerging from recent fairness literature highlights the dynamically self-undermining nature of group-based fairness constraints that include the outcome variable $Y$. \citet{Mishler2022} show that a classifier that is formally fair in the training distribution will violate the respective fairness constraint in the post-deployment distribution. For this reason, \citet{berk2021improving} argues for independence (demographic parity) as a fairness constraint, because it does not feature the outcome variable $Y$. \citet{Coston2020_CONF} suggests that the group-based fairness notion be formulated instead in terms of the potential outcomes $Y^d.$ These alternative proposals are no longer self-undermining, but they are still not probative of the policy's effect on structural inequality.  This paper's main contribution is to build upon the negative results of \citet{berk2021improving} and \citet{Mishler2022}: we show how the post-interventional effect of an algorithmically-informed policy on a structural inequality can be identified from a combination of (1) observational, pre-deployment data and (2) knowledge of the policy proposal. 

An emerging literature on long-term fairness focuses on the dynamic evolution of systems under sequential-decision making, static fairness constraints, and feedback loops; see \citet{Zhang2021_CONF} for a survey. \citet{Ensign2017} consider predictive feedback loops from selective data collection in predictive policing. \citet{Hu2018_CONF} propose short-term interventions in the labor market to achieve long-term objectives. Using two-stage models, \citet{Liu2019_CONF} and \citet{Kannan2019_CONF} show that procedural fairness constraints can, under some conditions, have negative effects on outcomes in disadvantaged groups. \citet{DAmour2020_CONF} confirm with simulation studies that imposing static fairness constraints does not guarantee that these constraints are met over time and can, under some conditions, exacerbate structural inequalities. Similar work is done by \citet{Zhang2020}. \citet{Creager2020_CONF} propose to unify dynamical fairness approaches in a causal DAG framework. Using time-lagged graphs, \citet{Hu2022} formulate a version of counterfactual long-term fairness. The picture emerging from this literature is that post-interventional outcomes of algorithmic policies are a relevant dimension for normative analysis that is not adequately captured by procedural fairness notions designed to hold in the training distribution. 

In this paper, we focus on using statistical profiling by public employment services to allocate the recently unemployed into active labor market programs. Respectively, \citet{Desiere2020} and \citet{Allhutter2020} provide detailed studies of the existing Flemish, and the proposed Austrian, algorithms. Using administrative data, \citet{Kern2021} perform a hypothetical analysis in a German setting. \citet{Scher2023} propose a dynamical model to study long-term and feedback effects on skills in a labor market context. To reduce inequality in the outcome distribution, \citet{Koertner2023} propose an inequality-averse objective function for the allocation of people into labor market programs. \citet{Kitagawa2019} and \citet{Viviano2023} make similar proposals in a more general setting. 

\section{Statistical profiling of the unemployed}\label{sec:statprofiling}
Since the 1990s, participation in active labor market programs (ALMPs) has been a condition for receiving unemployment benefits in many OECD countries \citep{Considine2017}. ALMPs take many forms, but paradigmatic examples include resume workshops, job-training programs and placement services, see \citet{Bonoli2010} for a helpful taxonomy. Evaluations of ALMPs across OECD countries find small but positive effects on labor market outcomes \citep{Card2018, Vooren2018, Lammers2019}. Importantly, the literature also reports large effect-size heterogeneity between programs and demographics, as well as assignment strategies that are as good as random for Switzerland \citep{Knaus2020a}, Belgium \citep{Cockx2023}, and Germany \citep{Goller2021}. This implies potential welfare gains from a more targeted allocation into programs, especially when taking into account opportunity costs--- a compelling motivation for algorithmic support.

Statistical profiling of the unemployed is current practice in various OECD countries including Australia, the Netherlands and Flanders, Belgium \citep{Desiere2019}. Paradigmatically, supervised learning techniques are employed to predict who is at risk of becoming long-term unemployed (LTU) \citep{Mueller2023_TECH_REPORT}. Such tools are regularly framed as introducing objectivity and effectiveness in the provision of public goods and align with demands for evidence-based policy and digitization in public administration. ALMPs target \textit{supply-side} problems by increasing human capital and \textit{matching} problems by supporting job search. \textit{Demand-side} policies that focus on the creation of jobs are not considered \citep{Green2022}.

Individual scores predicting the risk of long-term unemployment support a variety of decisions. For example, the public employment service (PES) of Flanders so far uses risk scores only to help caseworkers and line managers decide who to contact first, prioritizing those at higher risk \citep{Desiere2020}. In contrast, the PES of Austria (plans to) use risk scores to classify the recent unemployed into three groups: those with good prospects in the next six months; those with bad prospects in the next two years; and everyone else. The proposed policy of the Austrian PES is to focus support measures on the third group while offering only limited support to the other two. Advocates claim that, since ALMPs are expensive and would not significantly improve the re-employment probabilities of individuals with very good or very bad prospects, considerations of cost-effectiveness require a focus on those with middling prospects \citep{Allhutter2020}. However intuitive this may seem, it is nowhere substantively argued that statistical predictions of long-term unemployment from non-experimental data are reliable estimates for the effectiveness of labor-market programs. This is further complicated by the presence of long-standing structural inequalities in the labor market, which may be reproduced by algorithmic policies leaving those with ``poor prospects" to their own devices.
\begin{figure}[h]
  \centering
\includegraphics[scale=.35]{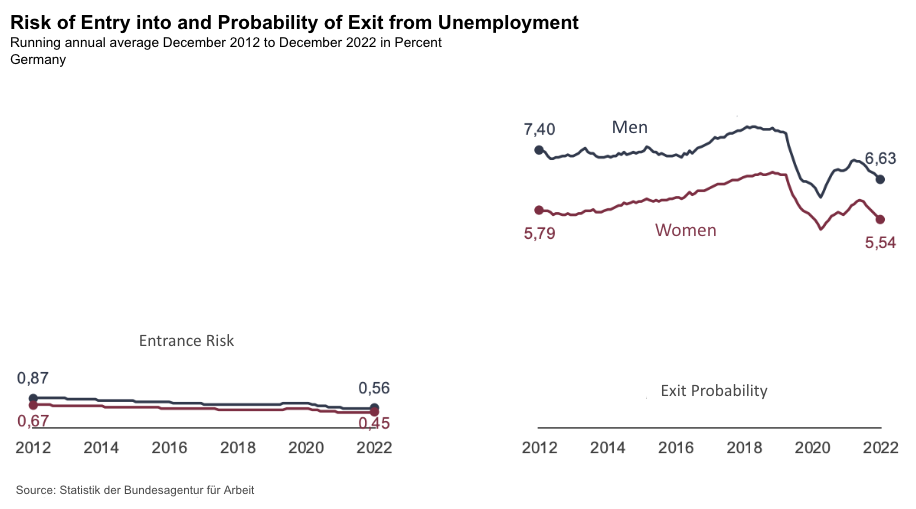}
   \caption{Data and Figure from the German PES \citep{arbeitsmarkt2022statistik}. The risk of entering unemployment is estimated as the number of newly registered unemployed divided by the number of employees subject to social insurance contributions. The exit probability from unemployment is estimated as the number of registered unemployed who find a job in the primary labor market relative to the number of registered unemployed. Both time series are running annual averages from December 2012 to December 2022.}
  \label{fig:entryexit}
\end{figure}

Indeed, labor markets in OECD countries are structured by various inequalities. Gender is a particularly long-standing and significant axis of inequality in labor markets, with the gender pay-gap and the child penalty being notorious examples \citep{Kleven2023, Bishu2016}. On the other hand, the gender gap in unemployment rates has largely disappeared over the last decades \citep{Albanesi2018}. Nevertheless, subtle structural differences in unemployment dynamics remain. For example, although women in Germany are less likely to enter into unemployment, their exit probabilities are also lower (see Figure \ref{fig:entryexit}). The obvious worry is that prediction algorithms will pick up on, entrench, or even exacerbate, these historical trends, as demonstrated in \citep{Kern2021}. The Austrian proposal for an LTU prediction algorithm furnishes a particularly dramatic example. That algorithm takes as input an explicitly gendered binary feature ``obligation to care'', which has a negative effect on the predicted re-employment probability and, by design, can be set to $1$ only for women \citep{Allhutter2020}. This controversial design choice was justified as reflecting the ``harsh reality" of the gendered distribution of care responsibilities. Whatever the wisdom of this particular variable definition, many other algorithms would pick up on the same historical patterns. Moreover, if the intended use of these predictions is to withhold support for individuals at high risk of long-term unemployment, it is clear that such a policy might exacerbate the situation by further punishing women for greater care obligations. Hopefully, the preceding motivates the need for a prospective fairness methodology that assesses whether women's re-employment probability suffers under a proposed algorithmic policy. More abstractly, what is needed is a way to predict how the pre-deployment probability $\Ppre(Y~|~A)$ will compare with the post-deployment probability $\Ppost(Y~|~A)$. With these estimates in hand, it would also be possible to predict whether the gender reemployment gap is exacerbated, or ameliorated, under a proposed algorithmic policy. The gender gap in reemployment probabilities is one particular choice for a fairness notion $\varphi(\cdot).$ Variations on this simple metric could be relevant in many other settings. For example, gender gaps in hiring, or racial disparities in incarceration could be criteria that an algorithmically informed policy should, minimally, not exacerbate. In the following section, we describe a methodology for predicting the evolution of reemployment probabilities from pre-deployment data.

\section{Performativity and Prospective Fairness}\label{sec:perfandpros}
First, some technicalities. Let $A,X,R,D,Y$ be discrete, {\em observed} random variables. In our example, $A$ represents gender; $X$ represents baseline covariates observed by the public employment service for the registered unemployed; $R$ is an estimated risk of becoming long-term unemployed; $D$ is an allocation decision made by the public employment service and $Y$ is a binary random variable that is equal to $1$ if an individual becomes long-term unemployed. For simplicity, we assume that $R$ is a deterministic function of $A$ and $X$. We write $\mathcal{A,X,R,D,Y}$ for the respective ranges of these random variables. For $d\in\mathcal{D},$ let $Y^d$ be the potential outcome under policy $d,$ in other words: $Y^d$ represents what the long-term unemployment status of an individual {\em would have been} if they had received allocation decision $d.$ Naturally, $Y^1,\ldots, Y^{|\mathcal D|}$ are not all observed. Our first assumption is a rather mild one; we require that the observed outcome for individuals allocated to $d$ is precisely $Y^d:$ 
\begin{equation} 
    \tag{\textsc{Consistency}} 
    Y = \sum_{d\in\mathcal{D}} Y^d\mathbbm{1}[D=d]. 
\end{equation}
Consistency is to be interpreted as holding both before and after the algorithmic policy is implemented.

More substantially, we assume that the potential outcomes and decisions are unconfounded given the observed features $(A, X)$ both before and after the intervention:
\begin{equation}
    \tag{\textsc{Unconfoundedness}}
    Y^d \independent_t D~|~A, X.
\end{equation}
Unconfoundedness is a rather strong assumption that requires that the observed features $A, X$ include all common causes of the decision and outcome. In the case of a fully automated algorithmic policy, unconfoundedness holds by design; but usually, risk assessment tools are employed to support human decisions, not fully automate them \citep{Levy2021}. Although it is not fated that all factors relevant to a human decision are available to the data analyst, unconfoundedness is reasonable if rich administrative data sets capture most of the information relevant to allocation decisions. For a case in which this assumption fails, see \citet{Petersen2021}.

We have argued that, in order to address the fundamental question of fair machine learning, one must predict whether implementing the candidate algorithmically-informed policy leads to an improvement, or at least no deterioration, in standards of justice. In the running example, this amounts to comparing features of $P_{\text{pre}}(Y~|~A)$ with $P_{\text{post}}(Y~|~A)$. The first distribution is trivial to estimate, but how to estimate $P_{\text{post}}(Y~|~A)$ from pre-deployment data? Here, the fundamental problem is performativity \citep{perdomo2020performative}. Our policy intervention will, in all likelihood, change the process of allocation into labor market programs and, thus, change the distribution of outcomes we are interested in. But not all kinds of performativity are equal. Some performative effects are intended and foreseeable. For example, the {\em algorithmic} effect is the intended change in decision-making due to algorithmic support:
\begin{equation*}
    P_{\text{pre}}\left(D=d~|~A=a, X=x\right) \neq P_{\text{post}}\left(D=d~|~A=a, X=x\right). \tag{\textsc{Algorithmic Effect}}
\end{equation*}
 Only the first term in this inequality is a quantity that can be directly estimated from training data. Nevertheless, it is possible to make reasonable conjectures about the second term given a concrete proposal for how risk scores should inform decisions. For example, if $D$ is binary, we could model the Austrian proposal as providing support so long as the risk score is neither too high nor low:
    \[ P_{\text{post}}(D=1~|~A=a, X=x) = \mathbbm{1} \left[l<R(a,x)<h\right].\]
More complex proposals for how risk scores should influence decisions will require more careful modelling. But careful modelling of the various ways in which predictions might influence decisions is precisely what we would like to encourage. 

Although we allow for algorithmic effects, these cannot be too strong---the policy cannot create allocation options that did not exist before. That is, the risk assessment tools only change allocation probabilities into {\em existing} programs. Moreover, we assume that the policy creates no unprecedented allocation-demographic combinations: 
\begin{equation}
    \tag{\textsc{No Unprecedented Decisions}}
    P_{\text{pre}}(D=d~|~A=a, X=x)>0 \text{ if } P_{\text{post}}(D=d~|~A=a, X=x)>0.
\end{equation}
This would be violated if e.g., no women were allocated to some program before the policy change.

Throughout this paper, we assume that no other forms of performativity occur. Some of these effects are neither intended nor to be expected. For example, we assume that the conditional average treatment effects (CATEs) of the allocation on the outcome are stable across time:
\begin{equation}
    \tag{\textsc{Stable CATE}} 
    P_{\text{pre}}\left(Y^d~|~A=a, X=x\right) = P_{\text{post}}\left(Y^d~|~A=a, X=x\right).
\end{equation}
This amounts to assuming that the effectiveness of the programs (for people with $A=a,X=x$) does not change, so long as all that has changed is the way we {\em allocate} people to programs.

While \textit{algorithmic effects} of deployment are intended and, to some degree, foreseeable types of performativity, \textit{feedback} effects that change the covariates are more complicated to model.\footnote{In the classification of \citet{Pagan2023_CONF}, we focus on what they call ``Outcome Feedback Loops''. In our terminology, performativity is not exhausted by feedback effects.} Following \citet{Mishler2022} and \citet{Coston2020_CONF}, we assume away the possibility of feedback effects, leaving these for future research:
\begin{equation}
    \tag{\textsc{No Feedback}}
    P_{\text{pre}}\left(A=a, X=x\right) = P_{\text{post}}\left(A=a, X=x\right).
\end{equation}
 \textsc{No Feedback} amounts to assuming that the baseline covariates of the recently employed are identically distributed pre- and post-deployment. Strictly speaking, this is false, since the decisions of caseworkers will affect the covariates of those who re-enter employment and some of them will, eventually, become unemployed again. However, since the pool of employed is much larger than the pool of unemployed, the policies of the employment service have much larger effects on the latter than the former. For this reason, we may hope that feedback effects are not too significant.

\textsc{No Unprecedented Decisions, Stable CATE and No Feedback} might fail dramatically if e.g., the deployment of the policy coincided with a major economic downturn. In a serious downturn, the employment service may have to assist people from previously stable industries (violating \textsc{No Unprecedented Decisions} and \textsc{No Feedback}), or employment prospects might deteriorate for everyone (violating \textsc{Stable CATE}). However, the possibility of such exogenous shocks is not a threat to our methodology. We are interested in the {\em ceteris paribus} effect of the algorithmic policy on structural inequality, not an all-thing-considered prediction of future economic conditions.

We are now in a position to show that, under the assumptions outlined above, it is possible to predict $P_{\text{post}}(Y=y~|~A=a)$ from pre-interventional data and a supposition about $P_{\text{post}}(D=d~|~A=a,X=x)$. That means that we can also predict changes to the overall reemployment probability $\Ppost(Y=0)$ as well as the gender reemployment gap $\Ppost(Y=y~|~A=1)-\Ppost(Y=y~|~A=0).$ Each of these are natural and important instances of $\varphi(\cdot).$ The proof is deferred to the supplementary material.

\begin{theorem}\label{theorem}
    Suppose that \textsc{Consistency, Unconfoundedness, No Unprecedented Decisions, Stable CATE} and \textsc{No Feedback} hold. Suppose also that $\Ppost(A=a)>0$. Then, $P_{\text{post}}(Y=y~|~A=a)$ is given by
\[ \sum_{(x,d)\in \Pi_{\text{post}}} P_{\text{pre}} (Y=y~|~A=a,X=x,D=d)P_{\text{pre}} (X=x~|~A=a)P_{\text{post}}(D=d~|~A=a,X=x),\]
where $\Pi_t = \left\{ (x,d) \in \mathcal{X \times D} : P_{t}(X=x,D=d~|~A=a)>0\right\}.$
\end{theorem}

Note that the first two terms in the product are identified from pre-deployment data. Given a sufficiently precise proposal for how risk scores influence decisions, it is also possible to model $\Pi_{\text{post}}$ and the last term before deployment. This allows us to systematically compare different (fairness-constrained) algorithms and decision procedures, and arrive at a reasonable prediction of their combined effect on reemployment probabilities (and the gender reemployment gap) before they are deployed. In the following, we show how this approach works in a toy model. However, in realistic high-dimensional settings, the first term might be estimated by regression and the second by multivariate density estimation. Finally, $\Ppost(Y=y|A=a)$ could be estimated by integration of the plug-in estimates.

\section{A Toy Model of a Public Employment Service}\label{sec:toymodel}
Our population of interest are the recently unemployed who have registered with some public employment service. For simplicity, we treat the gender variable $A$ as binary. Obligation to care $(X_1)$ is correlated with gender and increases the probability of long-term unemployment. In this model, the care-penalty is the only mechanism making gender a relevant axis of inequality. Educational attainment $(X_2)$ is independent of gender and increases the probability of finding a job. Prior to the deployment of statistical profiling, the assignment into a labor market program is modelled as random, with $40\%$ of the registered unemployed being allocated. This is consistent with empirical results by \citet{lechner2007value, Goller2021, Cockx2023}. These variables determine $Y_{\text{Prior}}$, a binary variable that is $1$ if the individual becomes unemployed in the long-term. High educational attainment, absence of care obligations, and participation in the labor market program all increase the reemployment probability.
\begin{align*}
    A \in \{0, 1\}      &\sim \text{Bernoulli}(0.5),              & 0:= \text{non-female;}\\
    X_1 \in \{0, 1\}    &\sim \text{Bernoulli}(0.2 + 0.4A),       & 0:= \text{no obligation to care;}\\
    X_2 \in \{0, 1\}    &\sim \text{Bernoulli}(0.2),              & 0:= \text{low educational attainment;} \\
    D_{\text{Prior}} \in \{0, 1\} &\sim \text{Bernoulli}(0.4),    & 0:= \text{no ALMP, and } \\
    Y_{\text{Prior}} \in \{0, 1\} &\sim \text{Bernoulli}(0.5 + 0.3X_1 - 0.2X_2 - 0.2D_{\text{Prior}})    &0:= \text{non-LTU.}
\end{align*}
Under the pre-deployment distribution, the gender reemployment gap is about $12$ percentage points, with $56\%$ of women and $44\%$ of non-women becoming long-term unemployed. The overall population probability of becoming long-term unemployed is $50\%.$ 

Although its budget only allows the employment service to allocate $40\%$ of the population to the program, it would like to make allocations more effective. To implement an algorithmic allocation policy, a logistic regression is trained on the features $A, X_1, X_2$ and the target variable $Y_{\text{Prior}}.$ The resulting risk score $R$ informs two potential policies, roughly resembling the Flemish policy of prioritizing the high-risk group and the Austrian policy of prioritizing the middle-risk group. Both policies fully automate allocation by thresholding the risk score. Under the Flemish-style policy, all and only individuals above the 60th risk percentile, $t_{F}$, are allocated into the program. Under the Austrian-style policy, the employment service restricts access to labor market programs to people above the 30th percentile $t_{\text{A-high}}$ and below the 70th percentile $t_{\text{A-low}}$. Due to sparse risk scores, the Austrian policy would allocate about $60\%$ of the population into programs. To ensure that the share of treated stays constant at $40\%$, we multiply the resulting assignment by a Bernoulli random variable $B$ parameterised by $\nicefrac{0.4}{0.6}=\nicefrac{2}{3}.$ All the assumptions of the previous section are satisfied by design; the example respects the causal structure of Figure \ref{fig:DAG_D0_D1}.
\begin{align*}
    B \in \{0, 1\}                  &\sim \text{Bernoulli}(\nicefrac{2}{3})                  & \\ 
    D_{\text{A}} \in \{0, 1\}       &= \mathbbm{1}[t_{\text{A-low}} \leq R \leq t_{\text{A-high}}] \times B     & 0:= \text{no ALMP; }\\
    Y_{\text{Post-A}} \in \{0, 1\}  &\sim \text{Bernoulli}(0.5 + 0.3X_1 - 0.2 X_2 - 0.2D_{\text{A}})                 & 0:= \text{non-LTU;} \\
    D_{\text{F}} \in \{0, 1\}       &= \mathbbm{1}[R \geq t_{F}]                                            & 0:= \text{no ALMP, and}\\
    Y_{\text{Post-F}} \in \{0, 1\}  &\sim \text{Bernoulli}(0.5 + 0.3X_1 - 0.2X_2 - 0.2D_{\text{F}})                  & 0:= \text{non-LTU.} 
\end{align*}
Neither the Flemish nor Austrian-style policies allocate anyone without care obligations and with high educational attainment $(X_1=0\wedge X_2=1)$ to the program. Focusing on those at high risk, the Flemish-style policy assigns all and only those with care obligations to the program, whether female or not. Since $60\%$ of women and $20\%$ of non-women have care obligations, this policy treats precisely $40\%$ of the population. Under the Austrian-style policy, women with low educational attainment but no care obligations $(X_1=0\wedge X_2=0)$ and those with care obligations but high educational attainment $(X_1=1 \wedge X_2=1)$ receive a 66\% chance of being allocated into the program; it denies the program to all other women. All others, except those $(X_1=0\wedge X_2=1)$, receive a 66\% chance of being allocated into the program.   

We would like to predict the overall reemployment probability, as well as the share of women and non-women that become long-term unemployed, after implementation. Analytically, we derive the following results for our toy model: the Flemish-style policy leaves the overall share of long-term unemployed unchanged (at $50\%$) while the Austrian-style policy slightly decreases long-term unemployment (to $49\%$). The Flemish-style policy brings the gender gap in long-term unemployment down from $12$ to $4$ percentage points by decreasing long-term unemployment among women $(P_{\text{post-F}}(Y=1~|~A=1)=52\%)$ and accepting an increased share ($48\%$) among the rest. The gender gap increases under the Austrian policy to $17$ percentage points. Under this algorithmic policy, women face higher long-term unemployment shares than before $(P_{\text{post-A}}(Y=1~|~A=1)=58\%)$, while the share among the others slightly decreases $(41\%)$. The detailed calculations are given in the Supplementary Material. Thus, it is possible to predict that the Austrian-style policy will exacerbate the gender reemployment gap, the Flemish-style policy will ameliorate it, and neither will have a large effect on the population reemployment probability. Since both policies rely on the same predictive model $R$, these differences would not be visible to internal fairness metrics.

\section{Conclusion and Future Work}\label{sec:Conclusion}
The deployment of an algorithmically informed policy is an intervention into the causal structure of society that can have important performative effects. Therefore, we argue for a prospective evaluation of risk assessment instruments: comparing the relevant structural inequalities at training time with the situation likely to arise {\em after} the algorithmic policies are deployed. If the algorithmic policy changes decision making, it is likely to change the distribution of the outcome variable. That undermines static, group-based fairness notions that include the outcome variable. But other retrospective fairness notions that are not self-undermining in this sense give no answer to the fundamental question of fair machine learning: whether the deployment of an algorithmic policy will exacerbate structural inequalities. We propose {\em prospective fairness} to explicitly respond to this fundamental worry motivating fair machine learning.

In this paper, we further develop a methodology for prospective fairness. We have shown that one can identify the effect of an algorithmic policy on a number of context-specific measures of structural inequality from the combination of (1) observational, pre-interventional data and (2) knowledge about the proposed policy. In our case study, the gender reemployment gap serves as a possible choice for such a measure $\varphi(\cdot)$. This result holds under a set of assumptions: \textsc{unconfoundedness} of the potential outcomes with the policy decisions, \textsc{stable conditional treatment effects}, \textsc{no feedback} effect onto the covariates, and \textsc{no unprecedented decisions}. We illustrate the proposal with a toy model of a public employment service. Two potential policies, one of prioritisation and one of efficiency, are informed by predictions of the risk of long-term unemployment. We show that it is possible to predict that in this setup the prioritisation policy will ameliorate the gender reemployment gap, while the efficiency policy will exacerbate it.

Future research should extend this work and its limitations. On a theoretical level, it is important to consider weaker assumptions to allow for the analysis of more complex situations. Most importantly, methods from dynamical causal modelling can be used to relax the \textsc{No Feedback} assumption. Furthermore, axiomatic approaches to the measurement of inequality from the theory of social choice may help narrow down the set of admissible fairness metrics $\varphi(\cdot)$ and elucidate the trade-offs between them. Our toy model can be extended to situations in which (1) the pre-interventional assignment is not random but informed by caseworkers' decisions; (2) the algorithm and caseworkers use different inputs; (3) risk scores only inform, but do not fully determine, the allocation decisions; and (4) allocation into the programme has heterogeneous treatment effects. Future work could also utilise this model set-up for a systematic comparison of the effect of different static fairness constraints on structural inequalities. In the future, we would like to apply this methodology to real administrative data from public employment services.

\acksection
This work has been funded by the Deutsche Forschungsgemeinschaft (DFG, German Research Foundation) under Germany’s Excellence Strategy – EXC number 2064/1 – Project number 390727645. The authors thank the International Max Planck Research School for Intelligent Systems (IMPRS-IS) for supporting Sebastian Zezulka.
\small
\bibliography{References}

\clearpage

\appendix
\section{Proof of Theorem~1}
\begin{proof}[Proof of Theorem~$1$]  \label{A:proof}

First, we need to show that all terms are well-defined. This amounts to showing that $P_\text{post}(A=a,X=x)$, $\Ppre(A=a)$ and $P_{\text{pre}}(A=a,X=x,D=d)$ are strictly greater than zero for all $(x,d)\in \Pi_{\text{post}}.$ 

We first show that $\Ppre(A=a)>0.$ Note that 

\begin{align*}
\Ppre(A=a) &= \sum_{x\in\mathcal{X}} \Ppre(A=a,X=x)\\
&= \sum_{x\in\mathcal{X}} \Ppost(A=a,X=x) \tag{\textsc{No Feedback}}\\
&= \Ppost(A=a)>0.
\end{align*}

We now show that $P_\text{post}(A=a,X=x)>0$ for all $(x,d)\in\Pi_{\text{post}}.$ Note that  

\begin{align*}
P_\text{post}(A=a,X=x) &= \Ppost(A=a)\sum_{e\in \mathcal{D}} P_\text{post}(X=x,D=e|A=a)\\
&\geq \Ppost(A=a) P_\text{post}(X=x,D=d|A=a)>0.
\end{align*}

Finally, we show that $P_{\text{pre}}(A=a,X=x,D=d)>0$ for all $(x,d)\in\Pi_{\text{post}}.$ Since $\Ppre(A=a)>0$, it suffices to show that $P_{\text{pre}}(X=x,D=d|A=a)>0$ for all $(x,d)\in\Pi_{\text{post}}.$ Accordingly, suppose that $(x,d)\in \Pi_{\text{post}}.$ Then 

\[P_{\text{post}}(X=x,D=d|A=a)=\Ppost(D=d|X=x,A=a)\Ppost(X=x|A=a)>0,\]

which entails that both $\Ppost(D=d|X=x,A=a)>0$ and $\Ppost(X=x|A=a)>0.$ By \textsc{No Unprecedented Decisions}, $\Ppre(D=x|X=x,A=a)>0$ and by \textsc{No Feedback} $\Ppre(X=x|A=a)>0.$ Therefore,

\[P_{\text{pre}}(X=x,D=d|A=a)=\Ppre(D=x|X=x,A=a)\Ppre(X=x|A=a)>0;\] 

and the question of well-definedness is settled.\\

Next, note that: $P_{\text{post}}(Y=y~|~A=a) =$

\begin{align*}
    &= \sum_{(x,d)\in\Pi_{\text{post}}} P_{\text{post}}(Y=y~|~A=a, X=x, D=d)P_{\text{post}}(X=x, D=d~|~A=a) \tag{\text{Total Probability}}  \\
    &= \sum_{(x,d)\in\Pi_{\text{post}}} P_{\text{post}}(Y=y~|~A=a, X=x, D=d)P_{\text{post}}(X=x|A=a)P_{\text{post}}(D=d~|~A=a, X=x)\\
    &= \sum_{(x,d)\in\Pi_{\text{post}}} P_{\text{post}}(Y=y~|~A=a, X=x, D=d)P_{\text{pre}}(X=x|A=a)P_{\text{post}}(D=d~|~A=a, X=x). \tag{\textsc{No Feedback}}
\end{align*}

Note that, whenever defined,

\begin{align*}
    P_{t}(Y=y~|~A=a, X=x, D=d) &= P_{t}\left(\sum_{e\in\mathcal{D}}Y^e\mathbbm{1}[D=e]=1~|~A=a, X=x, D=d\right) \tag{\textsc{Consistency}}  \\
    &=P_{t}\left(Y^d=y~|~A=a, X=x, D=d\right)  \\
    &=P_{t}\left(Y^d=y~|~A=a, X=x\right).   \tag{\textsc{Unconfoundedness}}
\end{align*}

\clearpage

Therefore,

\begin{align*}
    P_{\text{post}}(Y=y~|~A=a, X=x, D=d) &= P_{\text{post}}\left(Y^d=y~|~A=a, X=x\right)  \\
    &=P_{\text{pre}}\left(Y^d=y~|~A=a, X=x\right)   \tag{\textsc{Stable CATE}} \\
    &= P_{\text{pre}}(Y=y~|~A=a, X=x, D=d);
\end{align*}

and therefore $P_{\text{post}}(Y=y~|~A=a)=$ 

$$=\sum_{(x,d)\in\Pi_{\text{post}}} P_{\text{pre}}(Y=y~|~A=a, X=x, D=d)P_{\text{pre}}(X=x|A=a)P_{\text{post}}(D=d~|~A=a, X=x),$$
as required.
\end{proof}

\section{Analytical Computations for Toy Model} \label{A:analytical}

\begin{sidewaystable}
    \centering
    \caption{We tabulate the terms relevant for computing $P_\text{postA}(Y=1|A=1)$ and $P_\text{postF}(Y=1|A=1)$ according to Theorem 1. The eight column computes the products of terms in the third, fourth and fifth column. The ninth column computes the products of terms in the third, fourth and fifth column. Summing the terms in the eight column, we have that $P_\text{postA}(Y=1|A=1)=58\%.$ Summing the terms in the ninth column, we have that  $P_\text{postF}(Y=1|A=1)=52\%.$  }
    \begin{tabular}{*{10}{c}}
        \toprule
Care & Ed. & D & $\Ppre(Y=1|A,X,D=1,x,d)$ & $\Ppre(X=x|A=1)$ & $P_\text{postA}(D=d|A,X=1,x)$ & $P_\text{postF}(D=d|A,X=1,x)$ & &\\
\midrule
\midrule
0 & 0 & 0 & .5 & .64 & .334 & 1 & .107 & .32  \\
0 & 0 & 1 & .3 & .64 & .666 & 0 & .128 & 0  \\
0 & 1 & 0 & .3 & .16 & 1    & 1 & .048 & .048  \\
0 & 1 & 1 & .1 & .16 & 0    & 0 & 0 & 0  \\
1 & 0 & 0 & .8 & .16 & .334    & .043 & .384 & 0  \\
1 & 0 & 1 & .6 & .16 & .666    & .064 & 0 & .096 \\
1 & 1 & 0 & .6 & .04 & .334 & 0 & .008 & 0  \\
1 & 1 & 1 & .4 & .04 & .666 & 1 & .011 & .016 \\
    \bottomrule
    \end{tabular}

\bigskip\bigskip\bigskip

    \caption{We tabulate the terms relevant for computing $P_\text{postA}(Y=1|A=0)$ and $P_\text{postF}(Y=1|A=0)$ according to Theorem 1. The eight column computes the products of terms in the third, fourth and fifth column. The ninth column computes the products of terms in the third, fourth and fifth column.Summing the terms in the eight column, we have that $P_\text{postA}(Y=1|A=0)=41\%.$ Summing the terms in the ninth column, we have that  $P_\text{postF}(Y=1|A=0)=48\%.$  }
        \begin{tabular}{*{10}{c}}
            \toprule
Care & Ed. & D & $\Ppre(Y=1|A,X,D=0,x,d)$ & $\Ppre(X=x|A=0)$ & $P_\text{postA}(D=d|A,X=0,x)$ & $P_\text{postF}(D=d|A,X=0,x)$ & &\\
\midrule \midrule
0 & 0 & 0 & .5 & .32 & .334 & 1 & .053 & .16  \\
0 & 0 & 1 & .3 & .32 & .666 & 0 & .064 & 0  \\
0 & 1 & 0 & .3 & .08 & 1    & 1 & .024 & .024  \\
0 & 1 & 1 & .1 & .08 & 0    & 0 & 0 & 0  \\
1 & 0 & 0 & .8 & .48 & 1    & 0 & .384 & 0  \\
1 & 0 & 1 & .6 & .48 & 0    & 1 & 0 & .288 \\
1 & 1 & 0 & .6 & .12 & .334 & 0 & .024 & 0  \\
1 & 1 & 1 & .4 & .12 & .666 & 1 & .032 & .048 \\
            \bottomrule
    \end{tabular}
\end{sidewaystable}

\end{document}